\documentstyle[11pt,newpasp,twoside]{article}
\begin{document}
\title{The tidal perturbation of the low-mass globular cluster NGC\,5466}
\author{M. Odenkirchen, E.K. Grebel}
\affil{MPI f\"ur Astronomie, K\"onigstuhl 17, D-69117 Heidelberg, Germany}

\begin{abstract} 
The distribution of stars in the outer part of the sparse globular 
cluster NGC\,5466 provides evidence for possible tidal mass loss after 
a recent disk-shock near the solar circle.   
\end{abstract}

\section{Introduction}
NGC\,5466 is a metal-poor halo cluster located at $d = 16.3$~kpc from the 
Sun (Cassisi et al.\ 2001) and $z = 15.6$~kpc above the Galactic plane 
(see Fig.1). It is one of the less luminous clusters in the globular cluster 
system of our Galaxy.
It has a low central velocity dispersion of only 1.5~km\,s$^{-1}$ (Pryor 
\& Meylan 1993) and a relatively large core radius ($r_c = 6.4$~pc, Lehmann 
\& Scholz 1997). This implies a dynamical mass of about $5\times 10^4 
M_\odot$ (Pryor et al.\ 1991). 
The internal properties of NGC\,5466 are thus similar to - though less 
extreme than - those of the low-mass halo cluster Palomar\,5, which was 
recently found to have long tails of tidal debris (Odenkirchen et al.\ 2001, 
2003) that reveal ongoing tidal disruption.\\[1ex]
Existing measurements of the proper motion of NGC\,5466 
(see Odenkirchen et al.\ 1997) allow us to assess its local Galactic orbit.
We took the available kinematic data for NGC 5466 and a model of the Galactic 
potential developed by Allen \& Santillan (1991) and integrated the equations 
of motion to $\pm100$~Myr from present. The resulting orbit is plotted 
in Fig.1. It demonstrates that NGC\,5466 is on a highly eccentric orbit with 
$R_{max} > 40$~kpc, but has crossed the Galactic disk about 50~Myr ago at a 
distance of about 8~kpc from the Galactic center. 
This suggests that NGC\,5466 has recently undergone a tidal shock, which - 
due to the low mass and low concentration of this cluster - might 
lead to a significant loss of stars.

\section{Analysis of APM data}
We analysed the spatial distribution of stars in the outer parts of NGC\,5466 
using data from the APM catalogue\footnote{data available at 
www.ast.cam.ac.uk/\~{}apmcat/}. This catalogue is based on  
scans of the first Palomar Sky Survey (POSS-I) O and E plates. 
We extracted a $3^\circ \times 3^\circ$ field centered on the cluster and 
selected all sources classified as ``starlike''. 
To weed out stars that are most likely not associated with the cluster we 
restricted the sample to the color-magnitude box defined by $18.3 \le R \le 
20.0$ and $0.7 \le B-R \le 1.9$. 
This selects stars on the upper main sequence and near the main-sequence 
turn-off of NGC\,5466. Since the photometry obtained from the POSS-I plates 
is of moderate accuracy, the selection necessarily is rather coarse.\\[1ex]
Fig.2 shows the distribution of the selected cluster member candidates on 
the sky. While the inner part of the cluster appears spherical, there is 
a stellar halo around it, which has a non-spherical shape. 
This halo appears elongated towards the south-west and the north-east, i.e.,
stars are seen preferentially near the directions towards the Galactic center 
(see dashed line in Fig.1) and anticenter. The orientation of this cluster 
halo is further analysed in Fig.3, where we plot star counts in the range 
$9'\le r \le 18'$ as a function of position angle. The position angle 
of the major-axis of the cluster's halo is about +30~\deg while the position 
angle of the Galactic center-anticenter line is +43~\deg.  
This provides first evidence for a possible tidal perturbation of the cluster 
and eventual mass loss because the Galactic tidal field will drag 
stars away from the cluster along the center-anticenter line (see, e.g., the 
example of Pal\,5 in Odenkirchen et al.\ 2003). 
However, the APM data do not allow to trace cluster members to more than 
$20'$ from the center of NGC\,5466. Thus deeper and more accurate photometry 
is needed to explore the distribution of cluster stars in more detail.

\section{Comparison to simulated tidal debris}
Fig.4 shows a prediction of the spread of tidal debris from NGC\,5466.
We simulated tidal mass loss from the cluster by releasing particles
from the Lagrange points located on the Galactic centre - anticenter 
line in time steps of 10~Myr. The particles were assumed to have peculiar 
velocities of size 1.5~km\,s$^{-1}$ (equal to the measured velocity 
dispersion of the cluster). The dots show the present location 
of particles released $\le 200$~Myrs ago.
Stars lost during the last 50 Myrs, i.e., after the cluster's recent disk 
crossing, are highlighted as fat dots.
The distribution of the fat dots is similar to the observed halo of 
cluster member candidates around NGC\,5466. 
The orientation of the observed halo is compatible with a transition 
from the center-anticenter line towards alignment with the cluster's orbit.   
Deeper and cleaner data might allow to resolve the sharp bending 
near the Lagrange points suggested by the simulation.

\clearpage

\begin{figure}
\plotfiddle{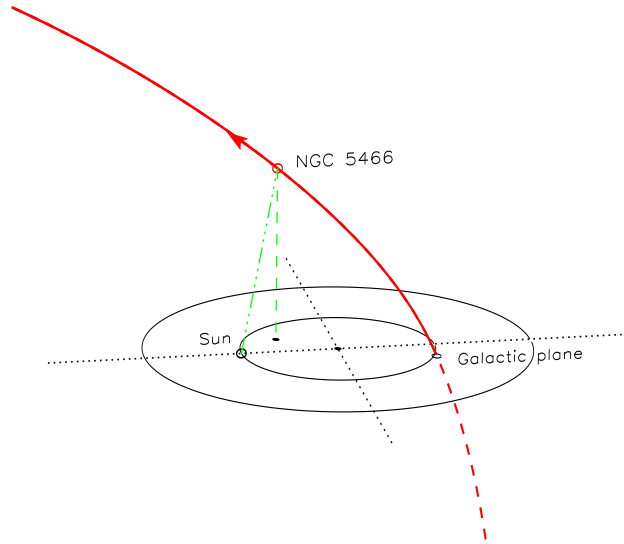}{7.0cm}{0}{60.0}{60.0}{-130}{-35}
\caption{The present location of NGC\,5466 and its Galactic orbit 
from $t = -100$~Myr to $t = +100$~Myr}
\end{figure}

\begin{figure}
\plotfiddle{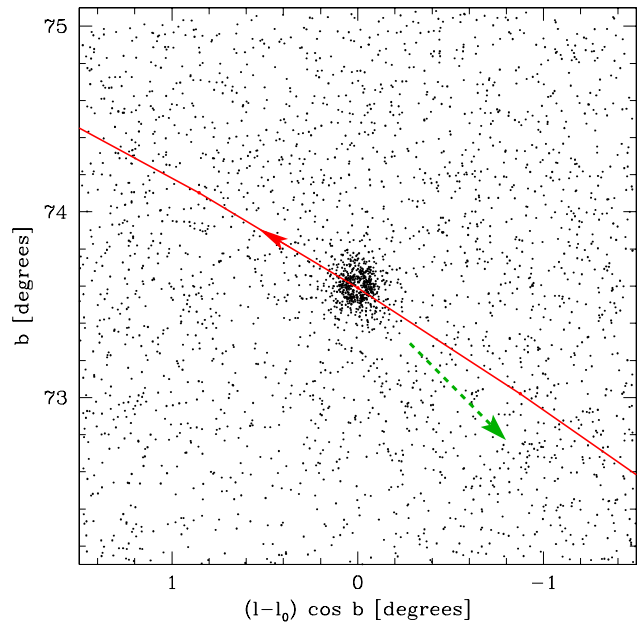}{7.0cm}{0}{60.0}{60.0}{-130}{-5}
\caption{Spatial distribution of cluster member candidates  
selected from the APM catalogue. Solid line: Orbit from Fig.1. 
Dashed line: direction to the Galactic center.}
\end{figure}

\begin{figure}
\plotfiddle{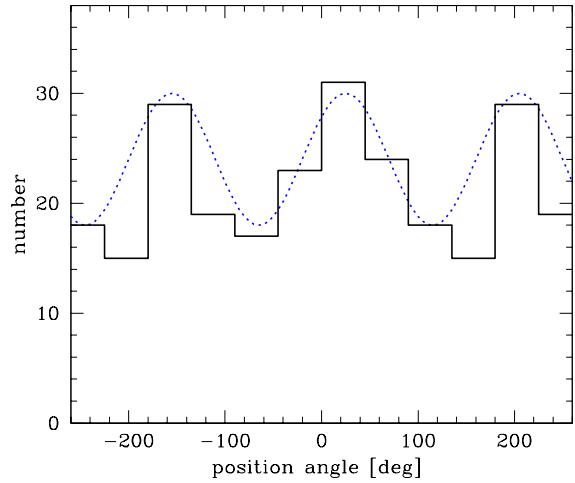}{7.0cm}{0}{60.0}{60.0}{-170}{-160}
\caption{Number of stars at $9'$ to $18'$ from the cluster center
counted in 45\deg\ sectors and plotted vs.\ position angle (from 
galactic north over east).}
\end{figure}

\begin{figure}
\plotfiddle{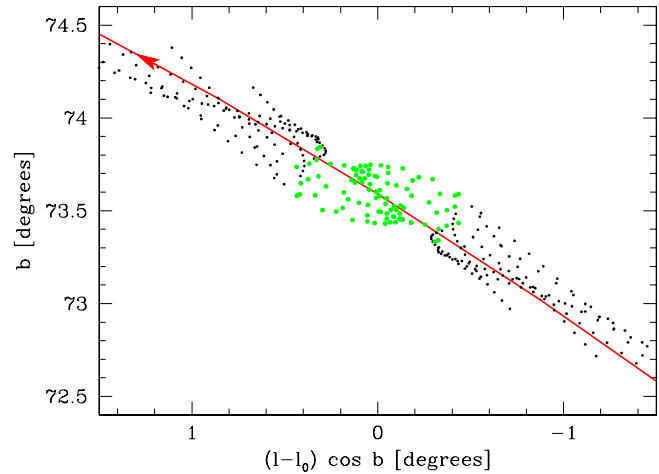}{7.0cm}{0}{60.0}{60.0}{-125}{0}
\caption{Simulated spread of tidal debris along the orbit of NGC\,5466. 
Dots show particles that were released from the cluster's inner and outer 
Lagrange point 
in time steps of 10~Myr with relative velocities of 1.5~km\,s$^{-1}$. 
Fat dots mark particles released after the last disk crossing 
(50~Myr ago). Note that the plot shows only unbound debris but not 
the main body of the cluster (center at $\Delta l \cos b =0\fdg0$, 
$b = 73\fdg6$).}
\end{figure}

\end{document}